\documentclass[aps,prl,showpacs,twocolumn,groupedaddress,floatfix,nofootinbib]{revtex4}
\usepackage{amsmath}
\usepackage{graphicx}
\usepackage{color}
\def\beq{\begin{eqnarray}}
\def\eeq{\end{eqnarray}}

\begin{document}

\title{The period of a classical oscillator}
\author{Paolo Amore}\email{paolo@cgic.ucol.mx} 
\author{Ricardo A. S\'aenz}\email{rasaenz@ucol.mx}
\affiliation{Facultad de Ciencias, Universidad de Colima, \\
Bernal D\'{\i}az del Castillo 340, Colima, Colima, Mexico} 

\date{\today}

\begin{abstract}    
We develop a simple method to obtain approximate analytical expressions 
for the period of a particle moving in a given potential. The method is 
inspired to the Linear Delta Expansion (LDE) and it is applied to
a large class of potentials. Precise formulas for the period are obtained.
\end{abstract}
\pacs{45.10.Db,04.25.-g}
\maketitle


In this letter we consider the problem of calculating the period of a unit mass 
moving in a potential $V(x)$. 
Although it is possible to solve this problem analytically only in a few cases, 
depending upon the form of the potential, several methods to find approximate 
results have been devised in the past. Many of the techniques that are used to 
solve this kind of problems are based on a perturbative expansion in some small 
parameter that appears in the equations of motion. 
This is the case of the Lindstedt-Poincar\'e method and of the multiple-scale method. 
Unfortunately the validity of these approaches is restricted to the domain of weak 
couplings and the series obtained rapidly diverge when larger couplings are considered. 
Recently, one of the authors and collaborators devised a non-perturbative version of the 
Lindstedt-Poincar\'e method, based on the ideas of the Linear Delta Expansion (LDE)\cite{lde}, 
which allows to obtain very accurate results in a wide class of non linear 
problems\cite{AA1:03,AA2:03,AM:04}.
In this letter we propose a different method, also inspired by the LDE, whose 
application is much simpler and for which convergence to the exact result can be proven.

Let us describe the method in detail. We consider a unit mass moving in a potential  
$V(x)$.  The total energy $E = \frac{\dot{x}^2}{2} + V(x)$ is conserved during the motion.
The exact period of the oscillations will be given by:
\begin{equation}
T = \int_{x_-}^{x_+} \frac{\sqrt{2}}{\sqrt{E-V(x)}}dx,
\label{period}
\end{equation}
where $x_\pm$ are the inversion points, obtained by solving the equation $E = V(x_{\pm})$.

Only in few cases it is possible to evaluate the integral (\ref{period}) analytically. 
In the spirit of the Linear Delta Expansion (LDE) we interpolate the full potential 
$V(x)$ with a solvable one $V_0(x)$\footnote{By solvable here we mean that the integral 
$\int_{x_-}^{x_+} \frac{\sqrt{2}}{\sqrt{E_0-V_0(x)}}dx$ can be done analytically.} 
defined as $V_\delta(x) = V_0(x) + \delta \ (V(x)-V_0(x))$.

We want to perform this interpolation without moving the inversion points;
for this reason we ask that $x_\pm$ be the inversion points also of the potential 
$V_0(x)$. As a result, the energy $E_0$ that the particle would possess if it was moving only 
in the potential $V_0(x)$ will be given by $E_0 = V_0(x_\pm)$.

We are now in position to rewrite eq.~(\ref{period}) as
\begin{multline}
T_\delta =\\ \int_{x_-}^{x_+} \frac{\sqrt{2}}{\sqrt{E_0-V_0(x)+\delta 
\big[ E -E_0 - V(x)+V_0(x)\big]}}  dx  .
\label{period1}
\end{multline}

We notice that Eq.~(\ref{period1}) reduces to Eq.~(\ref{period}) for $\delta=1$; 
for $\delta = 0$ this formula yields the period of oscillation between the points 
$x_\pm$ in the potential $V_0(x)$. We will treat the term proportional to 
$\delta$ as a perturbation and expand in powers of $\delta$. Since $V_0(x)$ depends 
upon one or more arbitrary parameters 
(which we will indicate with $\lambda$) a residual dependence upon these parameters 
shows up in the period when the expansion is carried out to a finite order. 
In order to eliminate such unnatural dependence we impose the Principle of 
Minimal Sensitivity (PMS)~\cite{Ste81} by requiring that
\begin{equation*}
\frac{\partial T}{\partial \lambda} = 0 . 
\end{equation*}

Finally, we can write explicitly the period by performing an expansion in $\delta$ 
and obtain
\begin{eqnarray}
T_\delta &=& \int_{x_-}^{x_+} \frac{\sqrt{2}}{\sqrt{E_0-V_0(x)}} \ 
\frac{dx}{\sqrt{1+\delta \ \Delta(x)}} \nonumber \\
         &=& \int_{x_-}^{x_+} \frac{\sqrt{2}}{\sqrt{E_0-V_0(x)}} \nonumber \\
         &&\times  \sum_{n=0}^\infty \frac{(2 n-1)!!}{n!   2^n} 
             (-1)^n  \delta^n  \left[\Delta(x)\right]^n  dx,
\label{period2a}
\end{eqnarray}
where 
\begin{eqnarray}
\Delta(x) = \frac{E -E_0 - V(x) + V_0(x)}{E_0-V_0(x)}.
\end{eqnarray}

$T_\delta$ can be written as
\begin{equation}\label{period2}
T_\delta = \sum_{n=0}^\infty \frac{(2 n-1)!!}{n!  \ 2^n} (-1)^n  \delta^n 
\int_{x_-}^{x_+} \frac{\sqrt{2} \ \big(\Delta(x)\big)^n}{\sqrt{E_0-V_0(x)}} dx
\end{equation}
provided that the series in Eq.~\eqref{period2a} converges uniformly, which is the case
if $|\Delta(x)|<1$ for every $x$, $x_- \le x \le x_+$.

We now apply Eq.~(\ref{period2}) to a few cases.  We start with the Duffing 
oscillator, which corresponds to the potential $V(x) = \frac{1}{2} \ x^2 + \frac{\mu}{4} \ x^4$.
Although the period of the Duffing oscillator can be calculated explicitly in terms 
of elliptic functions, we use this example to illustrate our method and 
to prove its efficiency.

We choose the interpolating potential to be $V_0(x) = \frac{1+\lambda^2}{2} \ x^2$ and obtain
\begin{equation*}
\Delta(x) =  \frac{2}{1+\lambda^2} \ 
\left[ \frac{\mu}{4} \ (A^2+x^2) - \frac{\lambda^2}{2} \right] \ . 
\label{duff2}
\end{equation*}

Hence the series in Eq.~(\ref{period2}) converges to the exact period for 
$\lambda > \lambda_0 \equiv \sqrt{\frac{\mu A^2}{2}} \ \sqrt{1-\frac{1}{\mu A^2}}$,
since $|\Delta(x)| <1$ uniformly for such values of $\lambda$ and $|x|\leq A$. 

The period of the Duffing oscillator calculated to first order using 
(\ref{period2}) is then
\begin{multline*}
T^{(0)}_\delta + \delta \ T^{(1)}_\delta =\\
\frac{2 \pi}{\sqrt{1+\lambda^2}} 
\left\{ 1 - \frac{\delta}{1+\lambda^2} \ \left[ \frac{3}{8}  \mu 
A^2-\frac{\lambda^2}{2}\right] 
\right\}
\label{duff3}
\end{multline*}

By setting $\delta=1$ and applying the PMS we obtain the optimal value of $\lambda$, 
$\lambda_{PMS} = \frac{\sqrt{3 \mu}}{2} A$, which remarkably coincides with the one
obtained in \cite{AA1:03} by using the LPLDE method to third order.
The period corresponding to the optimal $\lambda$ is 
\begin{equation*}
T_{PMS} = \frac{4  \pi}{\sqrt{4 + 3  \mu  A^2}} ,
\label{duff5}
\end{equation*}
which provides an error less than $2.2  \%$ to the exact period 
{\sl for any value of $\mu$ and $A$}.

It is useful to compare this result with the one in \cite{He03}, which differs from
our result only for a numerical factor under the square root. The result of \cite{He03} 
yields a much larger error ($16  \%$) over the period. This is expected, since our 
first order formula complies with the PMS.

\begin{figure}
\begin{center}
\includegraphics[width=7cm]{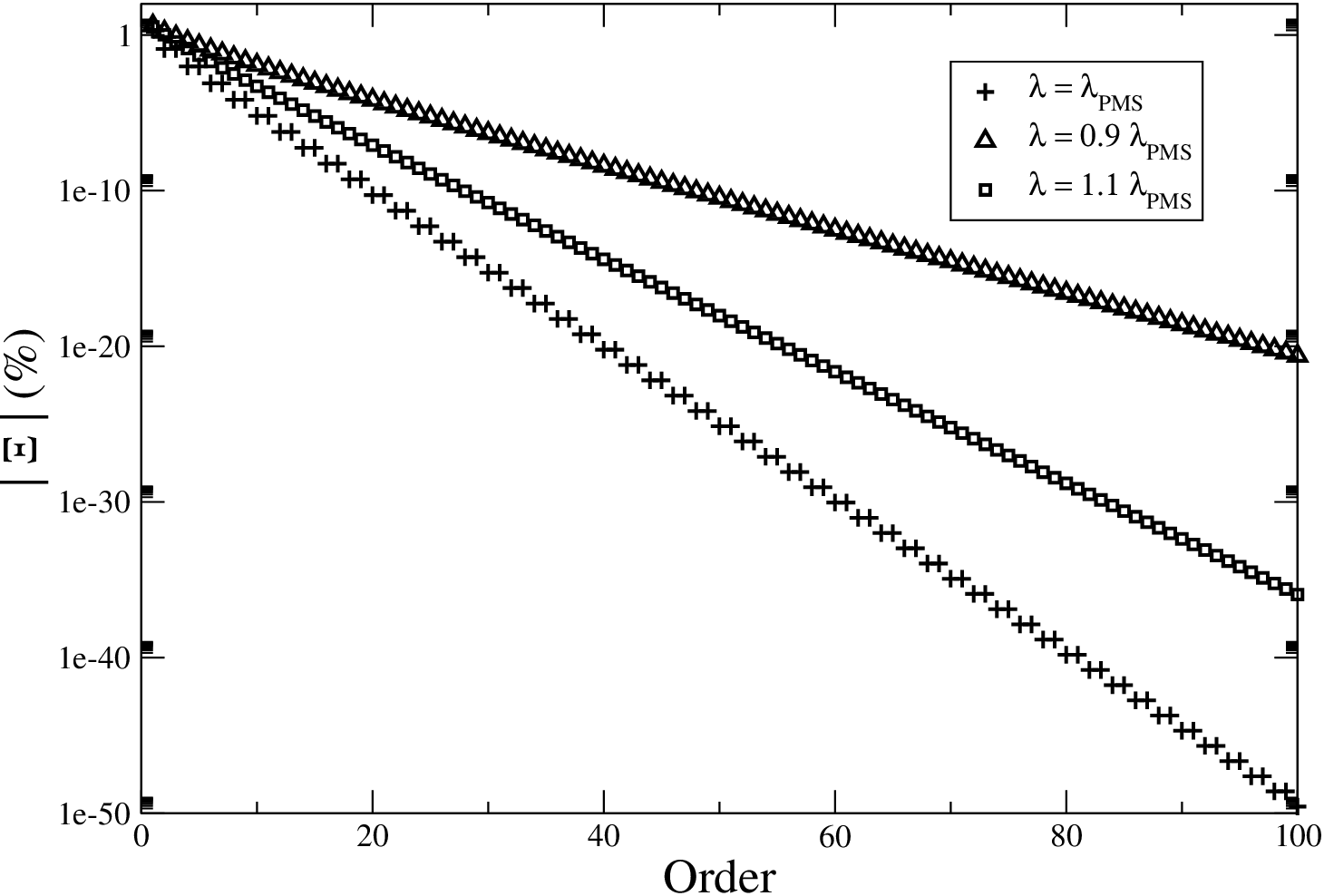}
\caption{Error over the period (in absolute value), defined as 
$\Xi \equiv \left[ \frac{T_{PMS}-T_{exact}}{T_{exact}} \right] \ 
\times 100$, for $A=10$ and $\mu=1$ as a function of the order. 
The three sets are obtained by using the optimal value 
$\lambda_{PMS} = \sqrt{3 \ \mu A^2}/2$ (plus), 
a value $\lambda = 0.9 \ \lambda_{PMS}$ (triangle) and 
$\lambda = 1.1 \ \lambda_{PMS}$ (square).}
\label{Fig_duff_conv1}
\end{center}
\end{figure}

Let us now come to the issue of convergence: since $\lambda_{PMS}>\lambda_0$ we can write
\begin{equation}
T_\delta = \sum_{n=0}^\infty  \delta^n  T_\delta^{(n)},
\label{duff_conv_3}
\end{equation}
where
\begin{multline}
T_\delta^{(n)} = \frac{(-1)^n\pi(2 n-1)!!}{2^{2 n-1}n!\sqrt{1+\lambda^2}}
\left( \frac{A^2  \mu - 2  \lambda^2}{1+\lambda^2} \right)^n \times\\
_2F_1\left(\frac{1}{2},-n,1,\frac{A^2\mu}{2  \lambda^2-A^2  \mu}\right) . 
\label{duff_conv_4}
\end{multline}
$_2F_1$ is the hypergeometric function. Since Eq.~\eqref{duff_conv_4} is 
essentially a power series, it converges exponentially to the exact result, which 
is precisely what we observe in  Figure~\ref{Fig_duff_conv1}, where  we plot the error 
$\Xi \equiv \left[ \frac{T_{PMS}-T_{exact}}{T_{exact}} \right] \ \times 100$ for 
three different values of the parameter $\lambda$ as a function of the order in the expansion. 
$T_{exact}$ is the exact period of the Duffing oscillator. Corresponding to the optimal value 
of the parameter,  $\lambda_{PMS} = \sqrt{3 \ \mu A^2}/2$, the rate of convergence is maximal.

We now consider the general anharmonic potential\footnote{The Duffing oscillator
considered in the previous example corresponds to choosing $N=2$ in the potential.} 
$V(x) = \frac{1}{2} \ x^2 + \rho \ \frac{x^{2 N}}{2 N}$ and obtain
\begin{equation*}
\Delta(x) =  \frac{2}{1+\lambda^2}
\left[ \frac{\rho}{2 N}\frac{A^{2 N}-x^{2 N}}{A^2-x^2} - \frac{\lambda^2}{2}\right].
\label{gao2}
\end{equation*}

To first order, Eq.~\eqref{period2} gives
\begin{multline*}
T^{(0)}_\delta + \delta   T^{(1)}_\delta =
\frac{2\pi}{\sqrt{1+\lambda^2}}\ \times\\
\bigg\{
1 + \frac{\delta}{1+\lambda^2}
\Big( - \rho \ A^{2 (N-1)}  \frac{\Gamma(N+1/2)}{\sqrt{\pi}  \Gamma(N+1)} + 
\frac{\lambda^2}{2} \Big) \bigg\},
\label{gao4}
\end{multline*}
in which case the optimal value of $\lambda$ is given by:
\begin{equation*}
\lambda_{PMS} = 
\sqrt{\frac{2  \rho  \Gamma(N+1/2)}{\sqrt{\pi}  \Gamma(N+1)}}  A^{N-1} .
\label{gao5}
\end{equation*}

This time, these values of $\lambda$ do not coincide with the ones obtained with 
the LPLDE method to third order 
for $N=3$ and $N=4$~\cite{AM:04}. With such $\lambda$ one obtains the expression
\begin{equation}
T_{PMS} = \frac{2 \pi}{\sqrt{1+ \displaystyle\frac{2 \rho A^{2 (N-1)} 
\Gamma(N+1/2)}{\sqrt{\pi} \Gamma(N+1)}}} .
\label{gao6}
\end{equation}

We have tested Eq.~(\ref{gao6}) for moderate values of $\rho$ and $A$ and seen that it provides 
a very good approximation to the exact period, even when 
the anharmonicity exponent $N$ gets very large.

We consider now the nonlinear pendulum, whose potential is given by
$V(\theta) = 1 - \cos\theta$.
By choosing the interpolating potential to be $V_0(\theta) = \frac{1+\lambda^2}{2} 
\theta^2$ we obtain
\begin{equation*}
\Delta(\theta) = - \frac{2}{1+\lambda^2} 
 \frac{\cos\Theta - \cos\theta}{\Theta^2-\theta^2} -1 ,
\end{equation*}
where $\Theta$ is the amplitude of the oscillations.
To first order our formula yields
\begin{equation*}
T_\delta = \frac{2\pi}{\sqrt{1+\lambda^2}} \Big( 1 + \frac{\delta}{2} \Big) - 
\frac{2  \pi  \delta}{(1+\lambda^2)^{3/2}}\frac{J_1(\Theta)}{\Theta},
\end{equation*}
where $J_1$ is the Bessel function of the first kind of order 1.
The optimal value of $\lambda$ in this case is given by
\begin{equation*}
\lambda_{PMS} = \sqrt{\frac{2J_1(\Theta)}{\Theta}-1}
\end{equation*}
and the period to first order is then
\begin{equation}
T_{PMS} = \pi\sqrt{\frac{2 \ \Theta}{J_1(\Theta)}}.
\label{Tpend1}
\end{equation}

Eq.~(\ref{Tpend1}) provides an excellent approximation to the exact period over
a wide range of amplitudes.

We now apply our expansion to two problems in General Relativity: 
the calculation of the deflection of the light by the Sun and the
calculation of the precession of a planet orbiting around the Sun. 
We use the notation of Weinberg~\cite{Weinberg}:
\begin{equation*}
B(r) = A^{-1}(r) = 1 - \frac{2 G  M}{r}.
\end{equation*}

The angle of deflection of the light by the Sun is given by the expression
\begin{equation*}
\Delta \phi = 2  \int_{r_0}^\infty  \sqrt{A(r)}  
\left[ \left(\frac{r}{r_0}\right)^2  \frac{B(r_0)}{B(r)} -1\right]^{-1/2} 
\frac{dr}{r} - \pi 
\end{equation*}
where $r_0$ is the closest approach.

With the change of variable $z = 1/r$ we obtain
\begin{equation*}
\Delta \phi = 2 \ r_0^{3/2}  \int_0^{1/r_0} 
 \frac{dz}{\sqrt{r_0 - r_0^3 z^2 - 2GM (1- r_0^3  z^2)}} - \pi,
\end{equation*}
which is exactly in the form required by our method. We introduce the
potential $V_0(z) = (r_0^3 - \lambda^2) \ z^2$ to obtain
\begin{equation*}
\Delta(z) = \frac{r_0 (1-r_0^2  z^2) - 2  G M \ (1 - r_0^3  z^3)}{(r_0^3 + \lambda^2)
\left( \frac{1}{r_0^2} - z^2\right)}.
\end{equation*}

By applying our method to first order we obtain 
\begin{multline}
\Delta\phi^{(0)}_\delta+\delta\Delta\phi^{(0)}_\delta = \\
-\pi+\frac{\pi\left(2+\displaystyle\frac{\left(2+\delta\right)  
\lambda^2}{r_0^3}\right)+
\displaystyle\frac{8\delta G M}{r_0}}%
{2{\left(1 + \displaystyle\frac{{\lambda}^2}{{{r_0}}^3}\right)}^{\frac{3}{2}}}.
\label{deflection1}
\end{multline}

The optimal value of the parameter $\lambda$ obtained by
the PMS is given by
\begin{equation*}
\lambda_{PMS} = 2  i  \sqrt{ \frac{2  G  M}{\pi}}  r_0.
\end{equation*}
The deflection angle corresponding to this value of $\lambda$ reads
\begin{equation}
\Delta\phi_{PMS} = - \pi + \sqrt{\frac{\pi}{1 - \displaystyle
\frac{8  G  M}{r_0  \pi}}}.
\label{deflectionPMS}
\end{equation}

The surface corresponding to the closest approach for which $\Delta\phi$ 
diverges is known as {\sl photon sphere} and for the Schwartzchild metric 
takes the value $r_0 = 3 G M$. 
It is remarkable that Eq.~(\ref{deflectionPMS}), despite its simplicity, 
is able to predict a slightly smaller {\sl photon sphere}, corresponding to 
$r_0 = 8 G M/\pi$. Clearly, this feature is completely missed 
in a perturbative approach.

\begin{figure}
\begin{center}
\includegraphics[width=7cm]{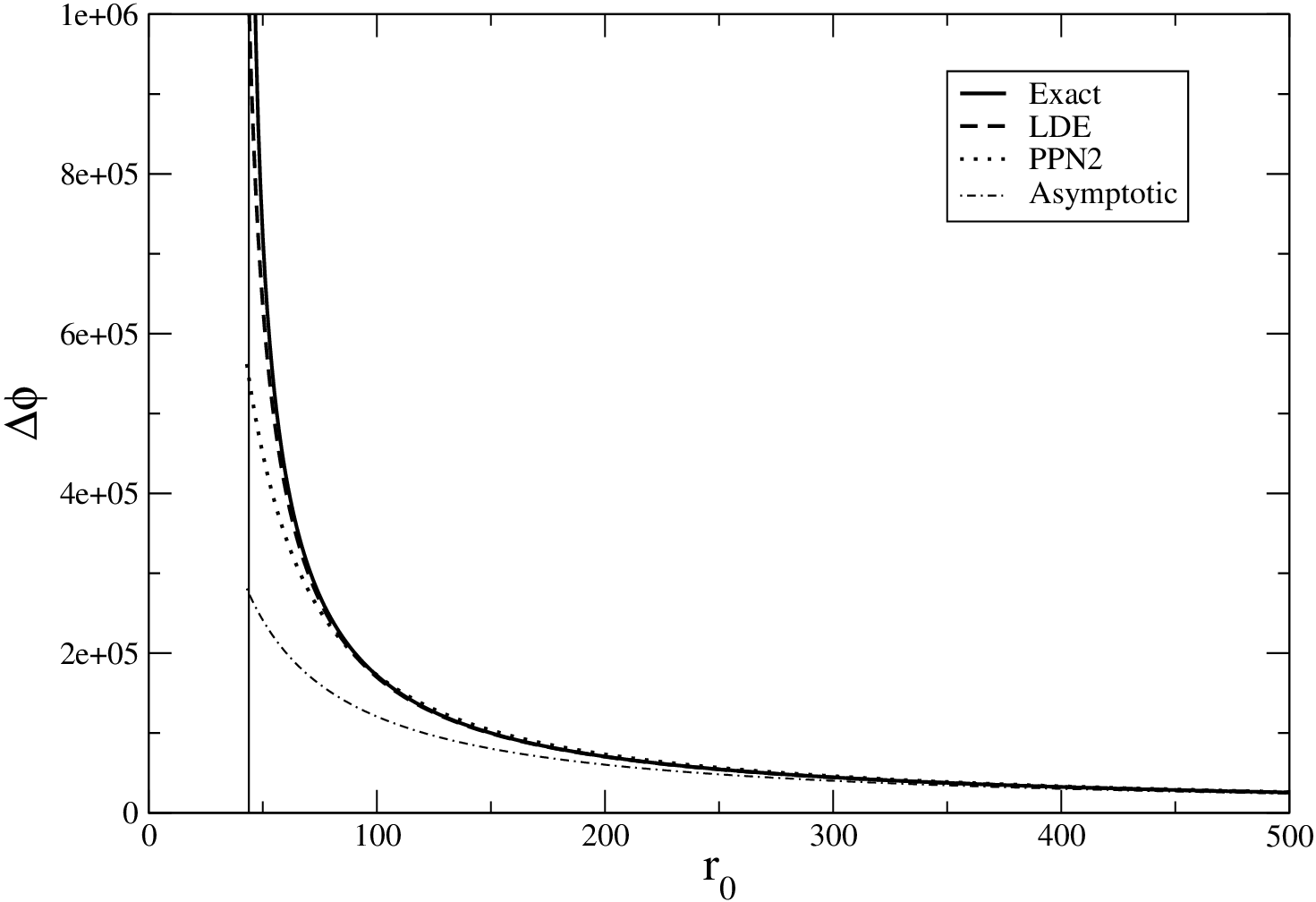}
\caption{Deflection angle of light obtained assuming 
$G/c^2 = 7.425 \times 10^{-30} \ m/kg$ and $M = 1.97 \times 10^{30} \ kg$ as 
function of the closest approach $r_0$. The solid line is the exact (numerical) 
result, the dashed line is obtained with Eq.~(\ref{deflectionPMS}), the dotted 
line is the post-post-Newtonian result of \cite{Epstein}, the dot-dashed line 
is the asymptotic result ($r_0 \rightarrow \infty$).
The vertical line marks the location of the photon sphere, where the deflection 
angle diverges.}
\label{Fig_deflection1}
\end{center}
\end{figure}

\begin{figure}
\begin{center}
\includegraphics[width=7cm]{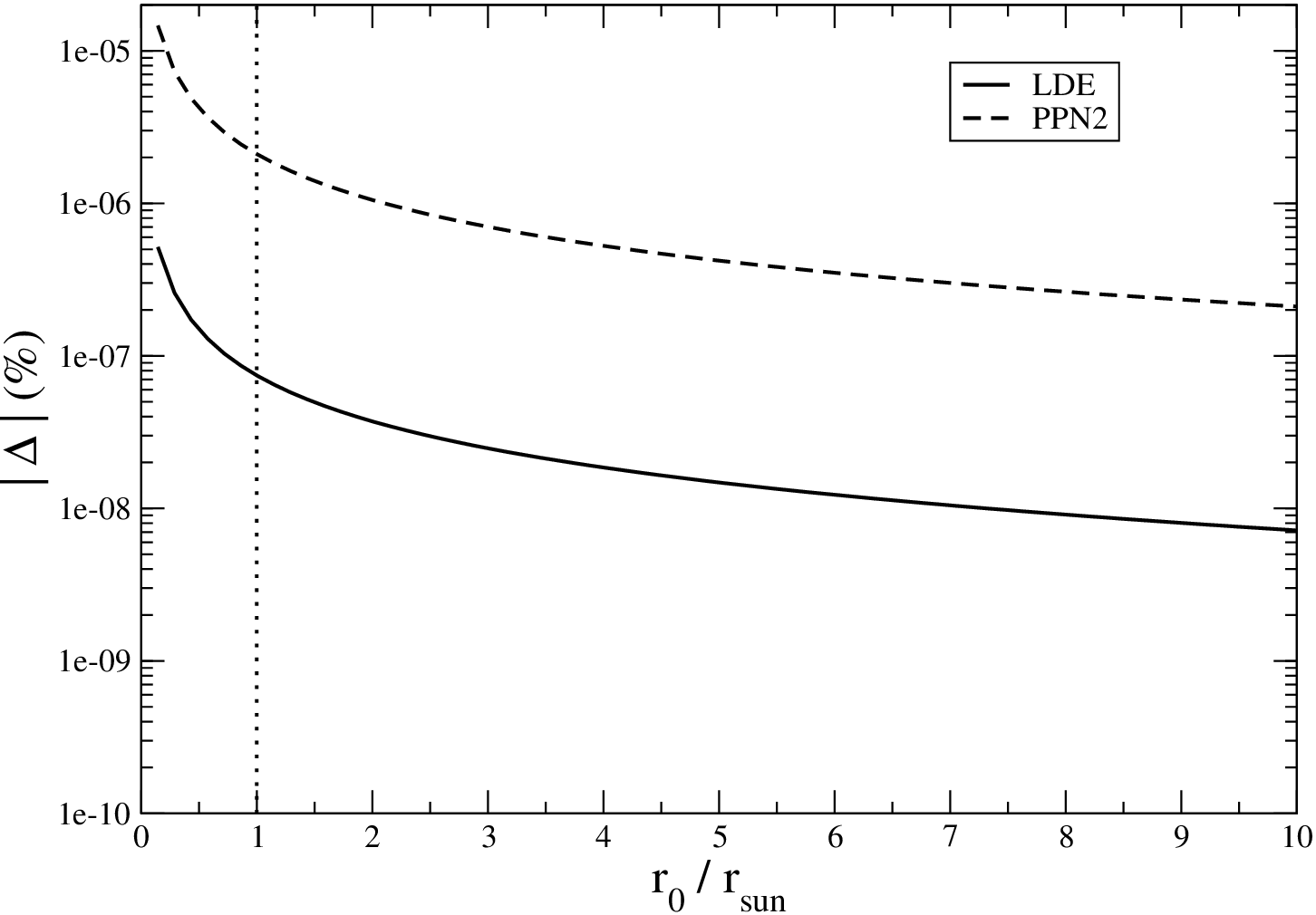}
\caption{Absolute value of the error over the deflection angle as a function of 
the closest approach in units of the sun radius 
($r_{sun} = 6.95 \times 10^{8} \ m$), assuming $G/c^2 = 7.425 \times 10^{-30} \ m/kg$ 
and $M = 1.97 \times 10^{30} \ kg$. The error is defined as 
$\Delta \equiv \frac{\Delta\phi_{approx}-\Delta\phi_{exact}}{\Delta\phi_{exact}}$.
The solid line is the error obtained with eq.~(\ref{deflectionPMS}),
while the dashed line is the error obtained using the PPN approximation of 
\cite{Epstein}.
The vertical line marks the solar radius.}
\label{Fig_deflection2}
\end{center}
\end{figure}

In Figure~\ref{Fig_deflection1} we compare Eq.~(\ref{deflectionPMS})
with the exact numerical result, the post-post-Newtonian (PPN) result 
of \cite{Epstein} and with the asymptotic result for very small
values of $r_0$ (close to the photon sphere). In Figure~\ref{Fig_deflection2} we plot
the error over the deflection angle obtained by using Eq.~(\ref{deflectionPMS}) 
or the PPN result in a range of $r_0$ of practical interest. Our formula gives an error
which is two orders of magnitude smaller than the one obtained with the PPN.

Finally we consider the problem of calculating the precession of the perihelion of a
planet orbiting around the Sun. The angular precession is given by~\cite{Weinberg}
\begin{equation}
\Delta \theta = -2  \pi - 2  \int_{r_-}^{r_+}
\frac{\sqrt{A(r)}\ dr}{r^2\sqrt{\displaystyle\frac{1}{J^2 B(r)} - \frac{E}{J^2} - 
\frac{1}{r^2}}}
\label{precession1}
\end{equation}
where
$E = \left(\frac{r_+^2}{B(r_+)}-\frac{r_-^2}{B(r_-)}\right)/\left(r_+^2-r_-^2\right)$
and $J^2 = \left(\frac{1}{B(r_+)}-\frac{1}{B(r_-)}\right)/\left(\frac{1}{r_+^2}-\frac{1}{r_-^2}\right)$.
$r_\pm$ are the shortest (perielia) and largest (afelia) distances from the sun. 
By the change of variable $z = 1/r$ we can write Eq.~(\ref{precession1}) as
\begin{multline*}
\Delta \theta = - 2  \int_{z_-}^{z_+}  \frac{1}{\sqrt{(z_+-z) (z-z_-)}}\times \\
\frac{dz}{\sqrt{(1- 2 G M (z+z_-+z_+))}} - 2  \pi,
\end{multline*}
where $z_\pm \equiv 1/r_\mp$.

We write 
\begin{multline*}
\Delta \theta_\delta =
 - 2\int_{z_-}^{z_+} \frac{1}{\sqrt{(z_+-z) (z-z_-)}}\times\\
\frac{dz}{\sqrt{(1-\lambda^2 + \delta( - 2 G M (z+z_-+z_+)+\lambda^2))}} - 2\pi. 
\end{multline*}

As usual we treat the term proportional to $\delta$ as a perturbation and expand 
to third order. The optimal 
value of $\lambda$ is obtained by using the PMS
\begin{equation*}
\lambda_{PMS} =  \sqrt{\frac{6 G M}{L}},
\end{equation*}
and yields the precession
\begin{multline}
\Delta \theta = \\
\frac{\pi(3 G^2 L M^2+ a ( -4 L^2 + 48 G L M - 147 G^2 M^2 ))}%
{4a (L - 6GM)^2 \sqrt{1 - \displaystyle\frac{6\ G\ M}{L}}},
\label{precession2}
\end{multline}
where $a$ is the semimajor axis of the ellipse, given by $a \equiv (r_-+r_+)/2$,
and $L$ is the {\sl semilatus rectum} of the ellipse, given by $1/L = (1/r_++1/r_-)/2$.

\begin{figure}
\begin{center}
\includegraphics[width=7cm]{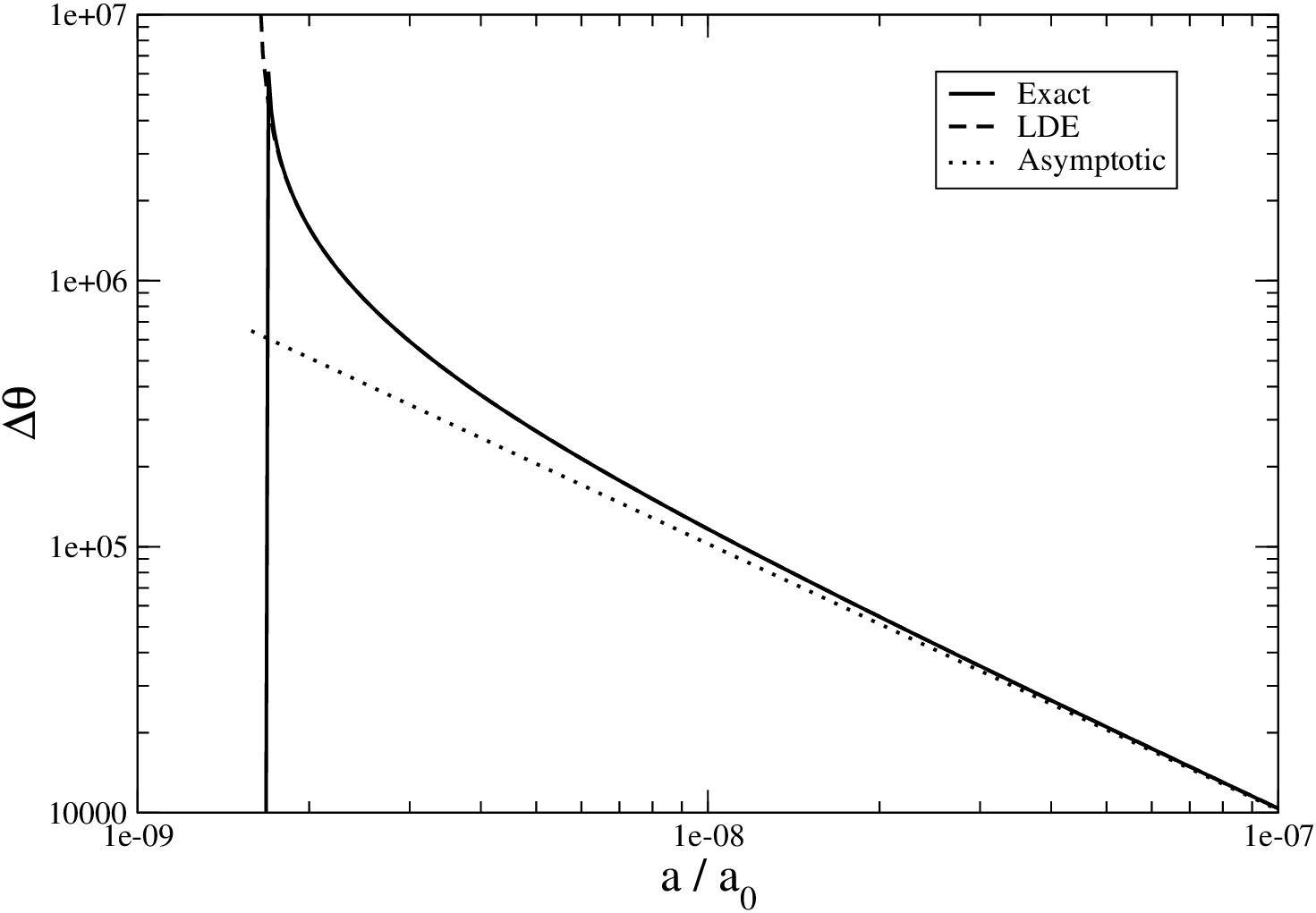}
\caption{Precession of the orbit of a planet assuming the values 
$M = 1.97 \times 10^{30} \ kg$, $G/c^2 = 7.425 \times 10^{-30} \ m/kg$ and
$\varepsilon = 0.2506$ (eccentricity). The scale of reference $a_0$ is taken to 
be the semimajor axis of Mercury's orbit ($a_0 = 5.971 \times 10^{10} \ 
m$). The solid line is the exact result, the dashed line is the result of 
Eq.~(\ref{precession2}) and the dotted line is the 
leading term in the perturbative expansion.}
\label{Fig_precession1}
\end{center}
\end{figure}

In Figure~\ref{Fig_precession1} we plot the precession of the orbit calculated 
through the exact formula (solid line), through Eq.~(\ref{precession2}) (dashed line) and 
through the leading order result~\cite{Weinberg} $\Delta\theta_0 = \frac{6\pi GM}{L}$ (dotted line) . Once again we find excellent 
agreement with the exact result.

In conclusion, we have devised a method to calculate with high accuracy a certain class of integrals, which
are very common in many physical problems. The convergence of our expansion to the exact result is easy to verify, as we
have explicitly shown in one special case.
Moreover the lowest order results obtained by applying our method already provide 
an excellent agreement with the full exact results. Work is currently in progress to apply this technique to 
a wider class of problems.

\begin{acknowledgments}
P.A. acknowledges support of Conacyt grant no. C01-40633/A-1.
\end{acknowledgments}

\end{document}